\documentclass[conference]{IEEEtran}
\IEEEoverridecommandlockouts
\usepackage{cite}
\usepackage{amsmath,amssymb,amsfonts}
\usepackage[nolist]{acronym}
\usepackage{hyperref}
\usepackage{scalerel}
\usepackage{graphicx}
\usepackage{nicefrac}
\usepackage{xcolor}
\usepackage{circuitikz}

\usetikzlibrary{calc}
\usepackage{pgfplots}
\usepgfplotslibrary{groupplots}

\pgfplotsset{compat=1.15}

\makeatletter
\let\MYcaption\@makecaption
\makeatother

\usepackage[font=footnotesize]{subcaption}

\makeatletter
\let\@makecaption\MYcaption
\makeatother

\DeclareMathOperator*{\argmin}{arg\,min}

\usetikzlibrary{positioning}
\usetikzlibrary{svg.path}

\def\BibTeX{{\rm B\kern-.05em{\sc i\kern-.025em b}\kern-.08em
    T\kern-.1667em\lower.7ex\hbox{E}\kern-.125emX}}

\definecolor{orcidlogocol}{HTML}{A6CE39}
\tikzset{
  orcidlogo/.pic={
    \fill[orcidlogocol] svg{M256,128c0,70.7-57.3,128-128,128C57.3,256,0,198.7,0,128C0,57.3,57.3,0,128,0C198.7,0,256,57.3,256,128z};
    \fill[white] svg{M86.3,186.2H70.9V79.1h15.4v48.4V186.2z}
                 svg{M108.9,79.1h41.6c39.6,0,57,28.3,57,53.6c0,27.5-21.5,53.6-56.8,53.6h-41.8V79.1z M124.3,172.4h24.5c34.9,0,42.9-26.5,42.9-39.7c0-21.5-13.7-39.7-43.7-39.7h-23.7V172.4z}
                 svg{M88.7,56.8c0,5.5-4.5,10.1-10.1,10.1c-5.6,0-10.1-4.6-10.1-10.1c0-5.6,4.5-10.1,10.1-10.1C84.2,46.7,88.7,51.3,88.7,56.8z};
  }
}

\newcommand\orcidicon[1]{\href{https://orcid.org/#1}{\mbox{\scalerel*{
\begin{tikzpicture}[yscale=-1,transform shape]
\pic{orcidlogo};
\end{tikzpicture}
}{|}}}}

\begin{acronym}
 \acro{R2M}{raw 2\textsuperscript{nd} moment}
 \acro{CSI}{channel state information}
 \acro{UE}{user equipment}
 \acro{EM}{electromagnetic}
 \acro{UL}{uplink}
 \acro{BS}{base station}
 \acro{TDD}{time division duplex}
 \acro{FDD}{frequency division duplex}
 \acro{ECC}{error-correcting code}
 \acro{MLD}{maximum likelihood decoding}
 \acro{HDD}{hard decision decoding}
 \acro{IF}{intermediate frequency}
 \acro{RF}{radio frequency}
 \acro{SDD}{soft decision decoding}
 \acro{NND}{neural network decoding}
 \acro{CNN}{convolutional neural network}
 \acro{ML}{maximum likelihood}
 \acro{GPU}{graphical processing unit}
 \acro{BP}{belief propagation}
 \acro{LTE}{Long Term Evolution}
 \acro{BER}{bit error rate}
 \acro{SNR}{signal-to-noise-ratio}
 \acro{ReLU}{rectified linear unit}
 \acro{BPSK}{binary phase shift keying}
 \acro{QPSK}{quadrature phase shift keying}
 \acro{AWGN}{additive white Gaussian noise}
 \acro{MSE}{mean squared error}
 \acro{LLR}{log-likelihood ratio}
 \acro{MAP}{maximum a posteriori}
 \acro{NVE}{normalized validation error}
 \acro{BCE}{binary cross-entropy}
 \acro{CE}{cross-entropy}
 \acro{BLER}{block error rate}
 \acro{SQR}{signal-to-quantisation-noise-ratio}
 \acro{MIMO}{multiple-input multiple-output}
 \acro{mMIMO}{massive multiple-input multiple-output}
 \acro{OFDM}{orthogonal frequency division multiplex}
 \acro{RF}{radio frequency}
 \acro{LoS}{line of sight}
 \acro{NLoS}{non-line of sight}
 \acro{NMSE}{normalized mean squared error}
 \acro{CFO}{carrier frequency offset}
 \acro{SFO}{sampling frequency offset}
 \acro{IPS}{indoor positioning system}
 \acro{TRIPS}{time-reversal IPS}
 \acro{RSSI}{received signal strength indicator}
 \acro{MIMO}{multiple-input multiple-output}
 \acro{ENoB}{effective number of bits}
 \acro{AGC}{automatic gain control}
 \acro{ADC}{analog to digital converter}
 \acro{ADCs}{analog to digital converters}
 \acro{FB}{front bandpass}
 \acro{FPGA}{field programmable gate array}
 \acro{JSDM}{Joint Spatial Division and Multiplexing}
 \acro{NN}{neural network}
 \acro{IF}{intermediate frequency}
 \acro{LoS}{line-of-sight}
 \acro{NLoS}{non-line-of-sight}
 \acro{DSP}{digital signal processing}
 \acro{AFE}{analog front end}
 \acro{SQNR}{signal-to-quantisation-noise-ratio}
 \acro{SINR}{signal-to-interference-noise-ratio}
 \acro{ENoB}{effective number of bits}
 \acro{PCB}{printed circuit board}
 \acro{EVM}{error vector mangnitude}
 \acro{CDF}{cumulative distribution function}
 \acro{MRC}{maximum ratio combining}
 \acro{MRP}{maximum ratio precoding}
 \acro{MRT}{maximum ratio transmission}
 \acro{DeepL}{deep-learning}
 \acro{DL}{downlink}
 \acro{SISO}{single-input single-output}
 \acro{SGD}{stochastic gradient descent}
 \acro{CP}{cyclic prefix}
 \acro{MISO}{Multiple Input Single Output}
 \acro{LMMSE}{linear minimum mean square error}
 \acro{ZF}{zero forcing}
 \acro{USRP}{universal software radio peripheral}
 \acro{RNN}{recurrent neural network}
 \acro{GRU}{gated recurrent unit}
 \acro{LSTM}{long short-term memory}
 \acro{NTM}{neural turing machine}
 \acro{DNC}{differentiable neural computer}
 \acro{TCN}{temporal convolutional network}
 \acro{FCL}{fully connected layer}
 \acro{MANN}{memory augmented neural network}
 \acro{RNN}{recurrent neural network}
 \acro{DNN}{deep neural network}
 \acro{FIR}{finite impulse response}
 \acro{BPTT}{back-propagation through time}
 \acro{GAN}{generative adversarial network}
 \acro{ELU}{exponential linear unit}
 \acro{tanh}{hyperbolic tangent}
 \acro{BICM}{bit-interleaved coded modulation}
 \acro{OTA}{over-the-air}
 \acro{IM}{intensity modulation}
 \acro{DD}{direct detection}
 \acro{RL}{reinforcement learning}
 \acro{SDR}{software-defined radio}
 \acro{WGAN}{Wasserstein generative adversarial network}
 \acro{BMD}{bit-metric decoding}
 \acro{BMI}{bit-wise mutual information}
 \acro{LDPC}{low-density parity-check}
 \acro{IDD}{iterative demapping and decoding}
 \acro{JSD}{Jensen-Shannon divergence}
 \acro{MMSE}{minimum mean square error}
 \acro{FFT}{fast Fourier transform}
 \acro{DFT}{discrete Fourier transform}
 \acro{IFFT}{inverse fast Fourier transform}
 \acro{QAM}{quadrature amplitude modulation}
 \acro{EMD}{earth mover's distance}
 \acro{TDL}{tapped delay line}
 \acro{KL}{Kullback-Leibler}
 \acro{PRACH}{physical random access channel}
 \acro{URLLC}{ultra-reliable low-latency communication}
 \acro{ANOMA}{asynchronous non-orthogonal multiple access}
 \acro{FEC}{forward error correction}
 \acro{PAPR}{peak-to-average power ratio}
 \acro{APP}{a posteriori probability}
 \acro{COTS}{commercial off-the-shelf}
 \acro{PLL}{phase locked loop}
 \acro{STO}{sampling time offset}
 \acro{SFO}{sampling frequency offset}
 \acro{CFO}{carrier frequency offset}
 \acro{CPO}{carrier phase offset}
 \acro{CSI}{channel state information}
 \acro{GNSS}{global navigation satellite system}
 \acro{ELAA}{extremely large aperture array}
 \acro{UE}{user equipment}
 \acro{DICHASUS}{\underline{Di}stributed \underline{Cha}nnel \underline{S}ounder by \underline{U}niversity of \underline{S}tuttgart}
 \acro{JCaS}{Joint Communication and Sensing}
 \acro{CT}{continuity}
 \acro{TW}{trustworthiness}
 \acro{KS}{Kruskal's stress}
 \acro{PCA}{principal component analysis}
 \acro{AoA}{angle of arrival}
 \acro{ToA}{time of arrival}
 \acro{TDoA}{time difference of arrival}
 \acro{ToT}{time of transmission}
 \acro{RD}{Rajski's distance}
 \acro{ADP}{angle-delay profile}
 \acro{MPC}{multipath component}
 \acro{CIR}{channel impulse response}
 \acro{MAE}{mean absolute error}
 \acro{MDS}{multidimensional scaling}
 \acro{t-SNE}{t-distributed stochastic neighbor embedding}
 \acro{SM}{Sammon's mapping}
 \acro{CIRA}{channel impulse response amplitude}
 \acro{CS}{cosine similarity}
 \acro{FCF}{forward charting function}
 \acro{CMD}{correlation matrix distance}
 \acro{UWB}{ultra-wideband}
 \acro{MUSIC}{MUltiple SIgnal Classification}
 \acro{FBCM}{forward-backward correlation matrix}
 \acro{MDL}{minimum descriptive length}
 \acro{BFGS}{Broyden-Fletcher-Goldfarb-Shanno}
 \acro{UPA}{uniform planar array}
 \acro{RMS}{root mean square}
 \acro{DRMS}{distance root mean square}
 \acro{CEP}{circular error probable}
 \acro{R95}{95\textsuperscript{th} percentile radius}
 \acro{eCDF}{empirical cumulative distribution function}
 \acro{L-LTF}{legacy long training field}
 \acro{HT-LTF}{high throughput long training field}
 \acro{DOI}{Digital Object Identifier}
 \acro{PA}{Power Amplifier}
 \acro{BSSID}{Basic Service Set Identification}
 \acro{RAN}{Radio Access Network}
\end{acronym}

\definecolor{mittelblau}{RGB}{0, 126, 198}
\definecolor{violettblau}{cmyk}{0.9, 0.6, 0, 0}
\definecolor{rot}{RGB}{238, 28 35}
\definecolor{apfelgruen}{RGB}{140, 198, 62}
\definecolor{gelb}{RGB}{1, 221, 0}
\definecolor{orange}{RGB}{244, 111, 33}
\definecolor{pink}{RGB}{237, 0, 140}
\definecolor{lila}{RGB}{128, 10, 145}
\definecolor{hellgrau}{RGB}{224, 224, 224}
\definecolor{mittelgrau}{RGB}{128, 128, 128}
\definecolor{dunkelgrau}{RGB}{80,80,80}
\definecolor{anthrazit}{RGB}{19, 31, 31}

\begin{document}

\title{Uncertainty-Aware Dimensionality Reduction for Channel Charting with Geodesic Loss
\thanks{This work is supported by the German Federal Ministry of Education and Research (BMBF) within the projects Open6GHub (grant no. 16KISK019) and KOMSENS-6G (grant no. 16KISK113).}}

\author{\IEEEauthorblockN{Florian Euchner\textsuperscript{\orcidicon{0000-0002-8090-1188}}, Phillip Stephan\textsuperscript{\orcidicon{0009-0007-4036-668X}}, Stephan ten Brink\textsuperscript{\orcidicon{0000-0003-1502-2571}} \\}
\IEEEauthorblockA{
Institute of Telecommunications, Pfaffenwaldring 47, University of  Stuttgart, 70569 Stuttgart, Germany \\ \{euchner,stephan,tenbrink\}@inue.uni-stuttgart.de
}
}

\maketitle

\begin{abstract}
Channel Charting is a dimensionality reduction technique that learns to reconstruct a low-dimensional, physically interpretable map of the radio environment by taking advantage of similarity relationships found in high-dimensional channel state information.
One particular family of Channel Charting methods relies on pseudo-distances between measured CSI datapoints, computed using dissimilarity metrics.
We suggest several techniques to improve the performance of dissimilarity metric-based Channel Charting.
For one, we address an issue related to a discrepancy between Euclidean distances and geodesic distances that occurs when applying dissimilarity metric-based Channel Charting to datasets with nonconvex low-dimensional structure.
Furthermore, we incorporate the uncertainty of dissimilarities into the learning process by modeling dissimilarities not as deterministic quantities, but as probability distributions.
Our framework facilitates the combination of multiple dissimilarity metrics in a consistent manner.
Additionally, latent space dynamics like constrained acceleration due to physical inertia are easily taken into account thanks to changes in the training procedure.
We demonstrate the achieved performance improvements for localization applications on a measured channel dataset.
\end{abstract}

\section{Introduction}
Wireless Channel Charting \cite{studer_cc} is a data-driven machine learning method that uses \ac{CSI} available at the base station to learn a map of relative \ac{UE} positions, known as the channel chart.
By leveraging the concept of dimensionality reduction, the channel chart is created without requiring labeled data, making Channel Charting a self-supervised technique.
The learned mapping from high-dimensional \ac{CSI} space to the low-dimensional latent space (the channel chart) is called \ac{FCF}.
Having access to an \ac{FCF} and hence relative locations enables various applications, such as absolute user localization after appropriate alignment of coordinate frames \cite{asilomar2023}, pilot allocation \cite{shaikh2024pilot}, or CSI compression for \ac{FDD} massive \ac{MIMO} \cite{csicompression}.
In the context of this work, we focus on Channel Charting in a distributed or non-distributed (massive) \ac{MIMO} system for user localization and evaluate our proposed technique in terms of localization performance metrics.
It is, however, reasonable to assume that proposals that improve localization accuracy will likely also have a positive impact on other applications of Channel Charting.

\begin{figure}
    \centering
    \begin{tikzpicture}
        \begin{groupplot}[
            group style={
                group size=2 by 1,
                x descriptions at=edge bottom,
                horizontal sep=0pt,
            },
            width=0.5\columnwidth,
            height=0.412\columnwidth,
            scale only axis,
            xmin=-15.5,
            xmax=6.1,
            ymin=-18.06,
            ymax=-1.5,
            ylabel shift = -8 pt,
            xlabel shift = -4 pt,
            xtick={-10, -6, -2, 2}
        ]
        \nextgroupplot[
            width=0.42\columnwidth,
            height=0.346\columnwidth,
            scale only axis,
            xlabel = {Coordinate $x_1 ~ [\mathrm{m}]$},
            ylabel = {Coordinate $x_2 ~ [\mathrm{m}]$},
            ylabel shift = -8 pt,
            xlabel shift = -4 pt,
            xtick={-10, -6, -2, 2}
        ]
            \addplot[thick,blue] graphics[xmin=-14.5,ymin=-17.06,xmax=4.1,ymax=-1.5] {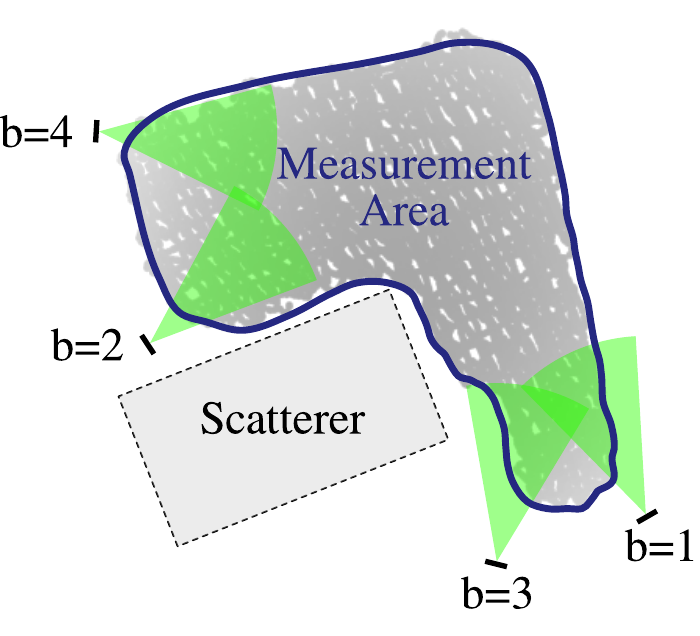};
    
        \nextgroupplot[
            width=0.42\columnwidth,
            height=0.346\columnwidth,
            scale only axis,
            xlabel = {Coordinate $x_1 ~ [\mathrm{m}]$},
            xlabel shift = -4 pt,
            xtick={-10, -6, -2, 2},
            ytick=\empty
        ]
            \addplot[thick,blue] graphics[xmin=-12.5,ymin=-14.5,xmax=2.5,ymax=-1.5] {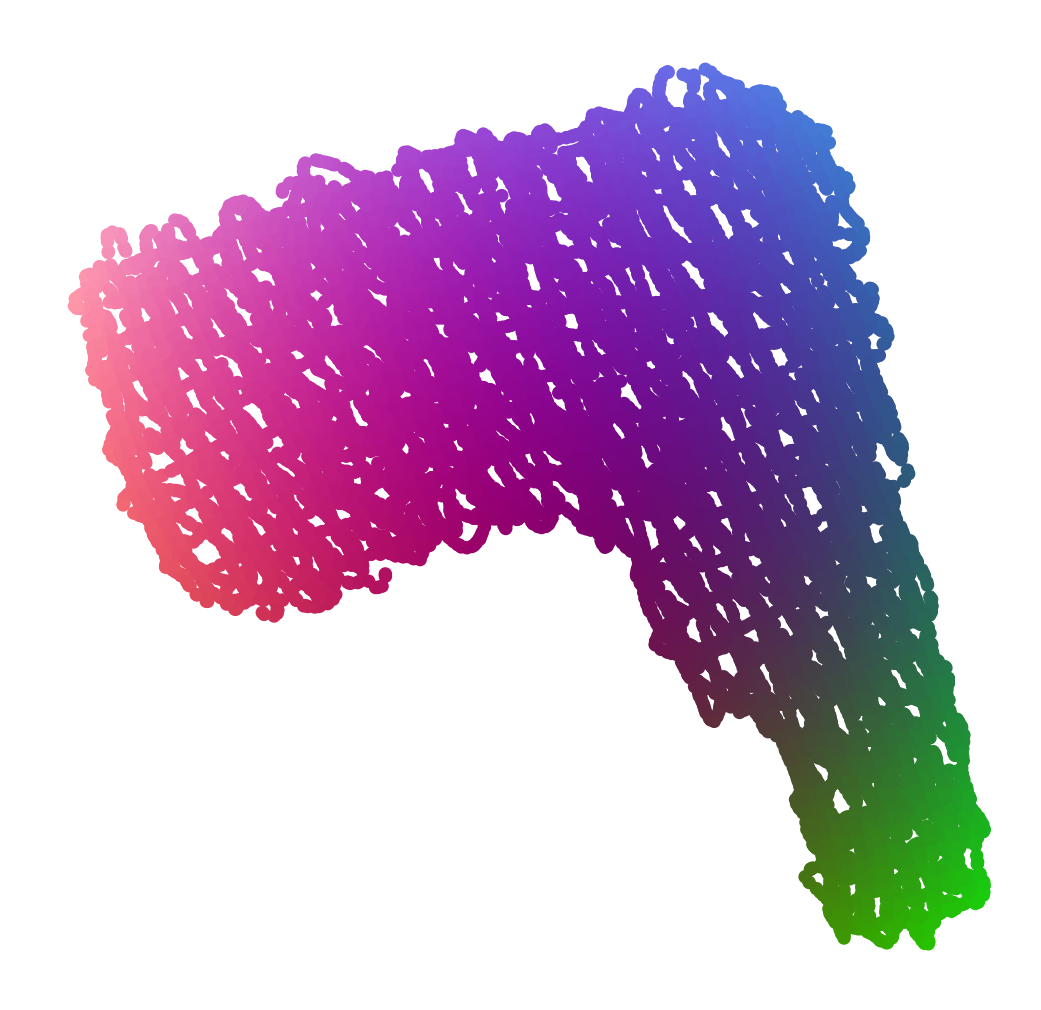};
        \end{groupplot}
    \end{tikzpicture}
    \vspace{-0.2cm}
    \caption{Information about dataset and environment: Top view map with antenna arrays drawn to scale, their orientation indicated by the green sectors (left) and scatter plot of colorized ``ground truth'' positions in $\mathcal S$ (right).}
    \vspace{-0.5cm}
    \label{fig:industrial_environment}
\end{figure}

Wireless localization systems or wireless communication systems with additional localization capabilities are already widely used.
For example, \acp{GNSS} solve the localization problem for most outdoor applications, \ac{UWB} enables fairly accurate positioning indoors and WiFi-based systems as well as cellular network-based systems provide approximate positions in cases where \ac{GNSS} or \ac{UWB} are unavailable or not feasible.
Regardless, the need for location or pseudo-location data is not always met by these existing systems:
For example, UE locations determined using a \ac{GNSS}-based system are usually not available to the \ac{RAN} of a cellular network and hence cannot be exploited for network optimization.
In other cases, devices may not integrate \ac{UWB} or \ac{GNSS} receivers for economic reasons, or systems may be unable to allocate additional spectral resources for localization.
Most importantly, most of these systems assume the existence of a \ac{LoS} channel, an assumption which often does not hold, leading to errors.
Channel Charting, on the other hand, does not make model assumptions, but exploits the uniqueness of multipath channels.
It relies on \ac{CSI}, which is usually available as a by-product of channel estimation for communication and hence does not need additional spectral resources or hardware.
Inference on the \ac{FCF} can be performed either on the \ac{UE} or the network side, wherever suitable \ac{CSI} is available, making Channel Charting ideal for applications that require access to pseudo-locations at the network side.

In this work, we revisit the idea of dissimilarity metric-based Channel Charting with a \ac{NN}-based \ac{FCF}, which has been developed and refined over the course of multiple contributions \cite{lei2019siamese, fraunhofer_cc, stephan2023angle}.
In particular, the results from \cite{asilomar2023} and \cite{stephan2023angle} are comparable to the results in this work, since they were obtained using the same real-world \ac{CSI} dataset, which is introduced in Sec.~\ref{sec:dataset}.
Two dissimilarity metrics from \cite{stephan2023angle}, which are also relevant in the context of this work, are briefly summarized in Sec.~\ref{sec:dissimilarity_metrics}.
In Sec.~\ref{sec:trainingarchitecture}, we propose a new batch-wise training architecture.
This architecture facilitates the improvements to the loss function that we propose in Sections~\ref{sec:acceleration},  \ref{sec:geodesic} and  \ref{sec:uncertainty}, each of which overcomes a deficiency of the existing methods.
In Sec.~\ref{sec:evaluation}, we evaluate the proposed changes and compare the performance to previous results.
Partial source code for this work that allows others to replicate our results has been made publicly available\footnote{See \url{https://github.com/Jeija/Geodesic-Uncertainty-Loss-ChannelCharting}}.

\section{Dataset and System Model}
\label{sec:dataset}
To be able to compare our results to previous work \cite{stephan2023angle, asilomar2023}, we evaluate the proposed changes to training architecture and loss function on the \emph{dichasus-cf0x} dataset \cite{dataset-dichasus-cf0x}.
The dataset was generated by our channel sounder \emph{\ac{DICHASUS}}, which is thoroughly described in \cite{dichasus2021}.
\ac{DICHASUS} measures the propagation channel between a single transmitter and many receive antennas and achieves long-term phase-coherence.
The considered dataset \emph{dichasus-cf0x} was captured in an industrial environment with $B = 4$ separate \acp{UPA} with half-wavelength antenna spacing, made up of $M_\mathrm{row} \times M_\mathrm{col} = 2 \times 4$ antennas each.
We represent the \ac{CSI} as time-domain \acp{CIR} $\mathbf H^{(l)} \in \mathbb C^{B \times M_\mathrm{row} \times M_\mathrm{col} \times T}$, where $T = 13$ denotes the number of taps along the time axis.
With a complex sampling rate of $50\,\mathrm{MSa}/\mathrm{s}$, $T = 13$ taps correspond to a captured \ac{CIR} duration of $260\,\mathrm{ns}$.
The single dipole transmit antenna is mounted on top of a robot, which travels along a set of trajectories inside a defined, L-shaped area, with an overall bounding box size of approximately $13\,\mathrm{m} \times 13\,\mathrm{m}$.
A tachymeter provides ``ground truth'' positions of the antenna $\mathbf x^{(l)}$ with millimeter-level precision and timestamps $t^{(l)}$ are also recorded.
Thus, the complete dataset can be represented as the following set containing a total of $L = 20851$ datapoints:
\[
    \text{Dataset}: \mathcal S = \left\{ \left(\mathbf H^{(l)}, \mathbf x^{(l)}, t^{(l)} \right) \right\}_{l = 1, \ldots, L}
\]
We assume the dataset is sorted in time, i.e., $t^{(l)} < t^{(l + 1)}$.
A top view map of the environment and the true datapoint positions $\mathbf x^{(l)}$ are shown in Fig.~\ref{fig:industrial_environment}.
A large metal container is located at the inner corner of the L-shape, blocking the \ac{LoS}.
The points have been colorized and the datapoints will retain their color even as the \acf{FCF} maps them to a position in the channel chart, which allows for a visual evaluation of the generated chart.
We define datasets $\mathcal S|_{b=\tilde b}$, $\tilde b = 1, \ldots, 4$, where
\[
    \mathcal S|_{b=\tilde b} = \left\{ \left(\mathbf H_{\tilde b}^{(l)}, \mathbf x^{(l)}, t^{(l)} \right) \Big\vert \left(\mathbf H^{(l)}, \mathbf x^{(l)}, t^{(l)} \right) \in \mathcal S \right\}.
\]
That is, the four datasets $\mathcal S|_{b=1}$, $\mathcal S|_{b=2}$, $\mathcal S|_{b=3}$ and $\mathcal S|_{b=4}$ are similar to $\mathcal S$, except that they only contain \ac{CSI} data from one out of the four antenna arrays.

\section{Dissimilarity Metrics}
\label{sec:dissimilarity_metrics}

Ever since Channel Charting was first proposed \cite{studer_cc}, various so-called dissimilarity metrics were investigated, which compute pseudo-distances between datapoints in the dataset.
Good dissimilarity metrics should not only be indicative of the true physical distance, but also be easy to compute, given that a dataset of cardinality $L$ contains $\frac{L^2-L}{2}$ distinct pairs of datapoints and may hence require up to this many dissimilarity metric computations.
In the following, we summarize the most promising dissimilarity metric approach found in \cite{stephan2023angle}.

\subsection{Timestamp Dissimilarity}
Datapoints which were measured in short succession of one another are likely also close in physical space.
This motivates the timestamp-based dissimilarity metric
\[
    \Delta_\mathrm{time}^{(i, j)} = \left\lvert t^{(i)} - t^{(j)} \right\rvert
\]
between the two datapoints with indices $i$ and $j$.
If $\Delta_\mathrm{time}$ is small, this is a good indicator of a small physical distance.
However, if $\Delta_\mathrm{time}$ is large, this may not necessarily imply a large distance between datapoints, since the \ac{UE} may also be static, move slowly or not move along a straight line.

\subsection{Angle Delay Profile Dissimilarity}
The \ac{ADP} of a channel at one antenna array describes how much power is received from which angle of arrival at which time delay according to the channel impulse response.
Intuitively, datapoints with similar \acp{ADP} are also likely to be close to each other in physical space.
The \ac{ADP} dissimilarity metric, first introduced in \cite{stephan2023angle}, formalizes this notion.
It first computes normalized powers $p_{b, \tau}^{(i, j)}$, indicating how well the datapoints' channels for array $b$ match at delay $\tau$, and then sums $1 - p_{b, \tau}^{(i, j)}$ over all arrays $b$ and channel taps $\tau$ to obtain a measure for the mismatch of the \acp{ADP} between datapoints $i$ and $j$:
\[
    \begin{split}
        \Delta_\mathrm{ADP}^{(i, j)} &= \sum_{b=1}^B \sum_{\tau=1}^{T} \left( 1 - p_{b,\tau}^{(i, j)} \right) \quad \text{with}\\
        p_{b,\tau}^{(i, j)} &= \frac{\left\lvert \sum\limits_{r=1}^{M_\mathrm{row}} \sum\limits_{c=1}^{M_\mathrm{col}} \left(\mathbf H_{b, r, c, \tau}^{(i)}\right)^* \mathbf H_{b, r, c, \tau}^{(j)}\right\rvert^2} { \left(\sum\limits_{r=1}^{M_\mathrm{row}} \sum\limits_{c=1}^{M_\mathrm{col}}\left\lvert \mathbf H_{b, r, c, \tau}^{(i)}\right\rvert^2 \right) \left(\sum\limits_{r=1}^{M_\mathrm{row}} \sum\limits_{c=1}^{M_\mathrm{col}} \left\lvert \mathbf H_{b, r, c, \tau}^{(j)}\right\rvert^2\right)}
    \end{split}
\]
The definition of $\Delta_{\mathrm{ADP}, i, j}$ is straightforwardly adapted to the single-array datasets $\mathcal S|_{b = \tilde b}$ by omitting the summation over $b$.
While $\Delta_{\mathrm{ADP}, i, j}$ may not be the best possible measure of \ac{CSI} dissimilarity, it can be computed efficiently, relying only on a closed-form expression made up of complex-valued multiplications, divisions and additions.

\section{Current and Proposed Training Architecture}
\label{sec:trainingarchitecture}
\begin{figure}
    \centering
    \includegraphics[width=0.42\columnwidth]{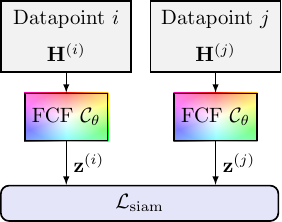}
    \vspace{-0.3cm}
    \caption{State of the art: Siamese neural network}
    \label{fig:siamese-nn}
    \vspace{-0.3cm}
\end{figure}

\begin{figure}
    \centering
    \includegraphics[width=0.7\columnwidth]{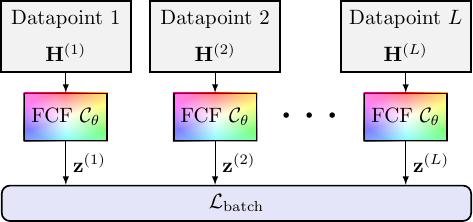}
    \vspace{-0.3cm}
    \caption{Proposed batch-wise loss training architecture}
    \label{fig:geodesic-nn}
    \vspace{-0.3cm}
\end{figure}

After the dissimilarity metrics $\Delta_\mathrm{ADP}^{(i, j)}$ and $\Delta_\mathrm{time}^{(i, j)}$ have been computed for all pairs $i$, $j$ in the dataset, they can be combined into a single dissimilarity metric $\Delta_\mathrm{fuse}^{(i, j)}$.
Furthermore, so-called geodesic dissimilarities $\Delta_\mathrm{geo}^{(i, j)}$ are obtained by running the shortest path algorithm on a $k$-neighbors graph of a full graph with $\Delta_\mathrm{fuse}^{(i, j)}$ as edge weights.
A more detailed explanation for these steps is provided in \cite{fraunhofer_cc, stephan2023angle} or in Sec.~\ref{sec:geodesic}, where the algorithm is revisited and an alternative is presented.

Since reference positions are assumed to be unavailable, the \ac{NN}-based \ac{FCF} $\mathcal C_\theta$ must be trained in a self-supervised manner.
For this, \cite{lei2019siamese} first proposed to use Siamese \acp{NN} in the context of Channel Charting, an approach that has has seen widespread adoption.
With a Siamese \ac{NN}, the loss function $\mathcal L_\mathrm{siam}$ is presented two predicted low-dimensional channel chart positions $\mathbf z^{(i)} = \mathcal C_\theta(\mathbf H^{(i)})$ and $\mathbf z^{(j)} = \mathcal C_\theta(\mathbf H^{(j)})$.
As illustrated in Fig.~\ref{fig:siamese-nn}, these two latent space representations $\mathbf z^{(i)}$ and $\mathbf z^{(j)}$ are computed in two forward passes of the time-domain \acp{CIR} through the \ac{FCF}, which can be interpreted as feeding $\mathbf H^{(i)}$ and $\mathbf H^{(j)}$ into two identical (i.e., weight-sharing) \acp{NN}, hence the name ``Siamese''.
Instead of comparing $\mathbf z^{(i)}$, $\mathbf z^{(j)}$ to the ground truth positions $\mathbf x^{(i)}$, $\mathbf x^{(j)}$, like a supervised training approach would, the Siamese loss function compares the distance $\left\lVert \mathbf z^{(i)} - \mathbf z^{(k)}\right\rVert$ to the geodesic dissimilarity $\Delta_\mathrm{geo}^{(i, j)}$.
For example,
\begin{equation}
    \mathcal L_\mathrm{siam} = \sum_{(i,j) \in \mathcal B} \frac{\left| \left\lVert \mathbf z^{(i)} - \mathbf z^{(j)} \right\rVert - \Delta_\mathrm{geo}^{(i, j)}\right|^2}{\Delta_\mathrm{geo}^{(i, j)} + \beta},
    \label{eq:siameseloss}
\end{equation}
has been used as a loss function.
Eq.~(\ref{eq:siameseloss}) is closely related to Sammon's loss \cite{sammon_mapping}, except for a neglected constant factor and the addition of $\beta$, which is a small value intended to improve training stability for small $\Delta_\mathrm{geo}^{(i, j)}$ and can be interpreted as the minimum uncertainty about the true distance.
The loss and the gradient of the loss function are computed over batches $\mathcal B$ of randomly selected datapoint pairs $(i,j)$.

The improvements to the loss function suggested in the following Sections~\ref{sec:acceleration}-\ref{sec:uncertainty} require a global view of the channel chart.
In other words, the loss function and hence also the weight update gradient is formulated in terms of the latent space representation $\mathbf z^{(l)}$ of many or all datapoints $l = 1, \ldots, L$, which requires all $\mathbf z^{(l)}$ to be available for loss computation.
This motivates what we call the batch-wise loss architecture illustrated in Fig.~\ref{fig:geodesic-nn}.
The idea of extending the loss function to take many samples into account instead of just two (Siamese \ac{NN}) or three (Triplet \ac{NN}) is inspired by work that demonstrated benefits of this scheme in a different context \cite{sohn2016improved}.
While it may seem as if batch-wise loss increases the computational complexity of training manyfold, this intuition is not correct for reasonable dataset sizes $L = |\mathcal S|$ when comparing to Siamese \ac{NN} training:
Training a good \ac{FCF} in a Siamese configuration requires large batches $\mathcal B$ of datapoint pairs $(i, j)$. For example, batch sizes of up to $|\mathcal B| = 6000$ pairs were used in \cite{stephan2023angle}, resulting in $2 \cdot 6000$ forward passes through the \ac{FCF}, which is on the same order of magnitude as just passing all $L = 20851$ datapoints in the dataset into the \ac{FCF} at once.
The loss function $\mathcal L_\mathrm{batch}$ for the new batch-wise training architecture may consist of multiple additive terms:
\begin{itemize}
    \item The Siamese loss $\mathcal L_\mathrm{siam}$ from Eq.~(\ref{eq:siameseloss}), which can be applied to the new architecture by generating batches $\mathcal B$ (of predefined size) of randomly selected datapoint pairs $(i, j)$ as part of the loss function.
    \item An acceleration contraint $\mathcal L_\mathrm{acc}$, described in Sec.~\ref{sec:acceleration}.
    \item A loss term $\mathcal L_\mathrm{geo}$, which is an alternative to $\mathcal L_\mathrm{siam}$ that uses geodesic distances in the channel chart as opposed to Euclidean distances, as described in Sec.~\ref{sec:geodesic}.
    \item A loss term $\mathcal L_\mathrm{geo,unc}$, which is an extension of $\mathcal L_\mathrm{geo}$ that also models uncertainties as defined in Sec.~\ref{sec:uncertainty}.
\end{itemize}
We would like to emphasize that $\mathcal L_\mathrm{siam}$, $\mathcal L_\mathrm{geo}$ and $\mathcal L_\mathrm{geo,unc}$ are mutually exclusive, while $\mathcal L_\mathrm{acc}$ can be added independently.

\section{First Improvement: Acceleration Constraint}
\label{sec:acceleration}
\subsection{Shortcoming of Current Approach}
Thanks to a dissimilarity metric that takes into account both \ac{CSI} and timestamps, we can ensure that datapoints which were measured in quick succession are also close to each other in space.
The laws of motion govern the movement of the \ac{UE} through physical space and its inertia and the force acting on the \ac{UE} dictate its acceleration.
Since the channel chart is a representation of the physical environment, the same laws should also apply there.
With the current Siamese \ac{NN} architecture, however, inertia is not taken into account at all, leading to unrealistic movement patterns in the channel chart.

\subsection{Suggested Solution}
While the true \ac{UE} movement patterns are unknown, some assumptions can be made.
For example, we assume that small (and zero) accelerations tend to occur more frequently than high accelerations and that there is a limit on the maximum acceleration.
As a simplification, we assume that $a = \left\lVert\mathbf a\right\rVert$ follows a folded normal distribution with location and scale parameters $\mu_\mathrm{acc}$ and $\sigma_\mathrm{acc}^2$ and suggest a negated log-likelihood acceleration constraint
\begin{equation}
	\mathcal L_\mathrm{acc} = -\frac{1}{L} \sum_{l = 1}^{L} \mathrm{ln} \left( \mathrm e^{-\frac{\left(\left\lVert \mathbf a^{(l)} \right\rVert - \mu_\mathrm{acc}\right)^2}{2 \sigma_\mathrm{acc}^2}} + \mathrm e^{-\frac{\left(\left\lVert \mathbf a^{(l)} \right\rVert + \mu_\mathrm{acc}\right)^2}{2 \sigma_\mathrm{acc}^2}} \right),
	\label{eq:acceleration-constraint}
\end{equation}
where $\mathbf v^{(l)} = \frac{\mathbf z^{(l)} - \mathbf z^{(l - 1)}}{t^{(l)} - t^{(l - 1)}}$ and $\mathbf a^{(l)} = \frac{\mathbf v^{(l)} - \mathbf v^{(l - 1)}}{t^{(l)} - t^{(l - 1)}}$ denote the instantaneous velocity and acceleration in latent space coordinates, i.e., in the channel chart.
We acknowledge that defining suitable parameters $\mu_\mathrm{acc}$ and $\sigma_\mathrm{acc}^2$ to approximate the true acceleration distribution may be challenging. 
When in doubt, using $\mu_\mathrm{acc} = 0$ and some large value $\sigma_\mathrm{acc}^2$ is a pragmatic choice, since that prohibits overly large accelerations without making too many assumptions.
If information about the true distribution of $a$ is available, $\mathcal L_\mathrm{acc}$ should be adapted accordingly instead of relying on Eq. (\ref{eq:acceleration-constraint}).

\section{Second Improvement: Geodesic Loss}
\label{sec:geodesic}
\subsection{Shortcoming of Current Approach}

\begin{figure}
    \centering
    \includegraphics[width=0.4\columnwidth]{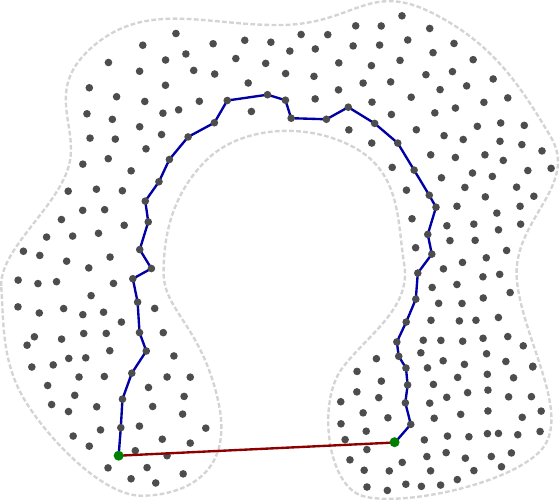}
    \caption{Illustration of locations of datapoints (gray dots) in nonconvex two-dimensional arrangement (in channel chart / physical space). The red line is the Euclidean distance between the two green datapoints, the blue line is one potential geodesic distance between the two datapoints.}
    \label{fig:nonconvex-sketch}
    \vspace{-0.3cm}
\end{figure}

\begin{figure}
    \centering
    \begin{tikzpicture}
        \node (start) [circle, below, minimum width = 0.3cm, minimum height = 0.3cm, fill=black!70!white] at (0, 0) {};
        \node (second) [circle, below, minimum width = 0.3cm, minimum height = 0.3cm, fill=black!70!white] at (2, 0) {};
        \node (third) [circle, below, minimum width = 0.3cm, minimum height = 0.3cm, fill=black!70!white] at (4, 0) {};
        \node (goal) [circle, below, minimum width = 0.3cm, minimum height = 0.3cm, fill=black!70!white] at (6, 0) {};

        \node[below=0.2cm of start] {$q^{(5, 8)}_0 = 5$};
        \node[below=0.2cm of second] {$q^{(5, 8)}_1 = 9$};
        \node[below=0.2cm of third] {$q^{(5, 8)}_2 = 7$};
        \node[below=0.2cm of goal] {$q^{(5, 8)}_3 = 8$};
        
        \draw [-latex, thick] (start) -- (second);
        \draw [-latex, thick] (second) -- (third);
        \draw [-latex, thick] (third) -- (goal);
    \end{tikzpicture}
    \vspace{-0.3cm}
    \caption{Illustration of $q^{(i, j)}_m$ path notation: Shortest path of length $M^{(i, j)} = 3$ from datapoint $i = 5$ to datapoint $j = 8$ is via datapoints $l=9$ and $l=7$.}
    \label{fig:path-illustration}
    \vspace{-0.3cm}
\end{figure}
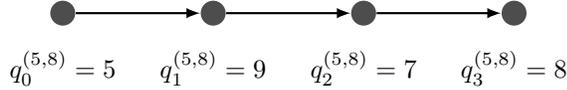

There is a crucial mismatch in the Siamese loss function in Eq.~(\ref{eq:siameseloss}), which has not been mentioned or addressed in Channel Charting literature so far:
The geodesic dissimilarity $\Delta_\mathrm{geo}^{(i,j)}$ is indicative of the \emph{geodesic distance} along the path from datapoint $i$ to datapoint $j$, whereas $\lVert \mathbf z^{(i)} - \mathbf z^{(j)} \rVert$ is the \emph{Euclidean distance} in latent space.
If the datapoints happen to be arranged in convex physical space, Euclidean and geodesic distances are similar and the mismatch is irrelevant.
However, if the physical arrangement of datapoints is nonconvex, $\mathcal L_\mathrm{siam}$ impermissibly compares Euclidean to geodesic distances, as illustrated in Fig.~\ref{fig:nonconvex-sketch}.
As a result, the learned channel chart is distorted.
This effect can be observed for the channel chart learned using loss function $\mathcal L_\mathrm{siam}$, shown in Fig.~\ref{fig:cc_allarrays_transformed}, where the $90^\circ$ angle in the corner of the L-shape is not preserved.

The described mismatch of geodesic and Euclidean distances is a well-known problem in dimensionality reduction and various solutions have been proposed \cite{budninskiy2019parallel,rosman2010nonlinear,zha2003isometric}, some even in the context of localization \cite{schwartz2019intrinsic}.
Our suggested solution attempts to solve this problem with the specific properties of Channel Charting in mind.

\subsection{Suggested Solution}
A better loss function without the explained mismatch is constructed by replacing the Euclidean distance $\lVert\mathbf z^{(i)} - \mathbf z^{(j)}\rVert$ with a geodesic distance term that takes into account the path that was taken when computing $\Delta_\mathrm{geo}^{(i, j)}$.
We revisit how $\Delta_\mathrm{geo}^{(i, j)}$ is calculated from $\Delta_\mathrm{fuse}^{(i, j)}$ (for any combination $i, j$) to explain the differences of our approach:
\begin{itemize}
    \item First, we construct a full weighted graph $G$. Each vertex in the graph corresponds to one datapoint $l = 1, \ldots, L$. We assign weight $\Delta_\mathrm{fuse}^{(i, j)}$ to the edge between vertices $i$ and $j$.
    \item Second, for every datapoint $l$, find the $k$ nearest neighbors (i.e., connected edges with smallest weights), and remove all other edges, forming a $k$-neighbors graph $\tilde G$. Typically, values around $k = 20$ are chosen.
    \item Third, $\Delta_\mathrm{geo}^{(i, j)}$ is the length of the shortest (lowest-weight) path between vertices $i$ and $j$ found by a shortest path algorithm applied to $\tilde G$.
\end{itemize}

Our proposed solution extends the third step:
As an additional output of the shortest path algorithm, we remember the path taken to get from datapoint $i$ to datapoint $j$ for any $i,j$.
We denote the index of the $m$\textsuperscript{th} datapoint along the path from $i$ to $j$ as $q^{(i, j)}_m$, and the number of hops in the path from $i$ to $j$ as $M^{(i, j)}$.
Hence, $q^{(i, j)}_0 = i$ and $q^{(i, j)}_{M^{(i, j)}} = j$, as illustrated in Fig.~\ref{fig:path-illustration}.
In the practical implementation, we do not store paths $q^{(i, j)}$ directly, but instead store a so-called predecessor matrix of size $L \times L$ and construct the paths on-the-fly during training from that matrix to conserve memory.
Therefore, the amount of memory required for storing all shortest paths through the graph scales with $L^2$, which is on the same order as the memory that is required for storing all values of $\Delta_\mathrm{geo}^{(i, j)}$, so there is no significant increase in terms of the required memory.
For notational reasons, however, we resort to the path notation $q^{(i, j)}_m$ in the following paragraphs, as reconstructing paths from the predecessor matrix is a straightforward procedure anyway.
We find the (geodesic) length $\rho_\mathrm{geo}^{(i,j)}$ of the path $q^{(i,j)}_m$ through the channel chart to be
\[
    \rho_\mathrm{geo}^{(i, j)} = \sum_{m=1}^{M^{(i, j)}} \left\lVert\mathbf z^{\left(q^{(i, j)}_{m-1}\right)} - \mathbf z^{\left(q^{(i, j)}_m\right)} \right\rVert,
\]
where, $\mathbf z^{(l)} = \mathcal C_\theta\left(\mathbf H^{(l)}\right)$ denotes the latent space representation for datapoint $l$ as usual.
We define the geodesic loss function $\mathcal L_\mathrm{geo}$ by replacing the Euclidean distance in Eq.~\ref{eq:siameseloss} by $\rho_\mathrm{geo}$:
\begin{equation}
    \mathcal L_\mathrm{geo} = \sum_{(i,j) \in \mathcal B} \frac{\left| \rho_\mathrm{geo}^{(i, j)} - \Delta_\mathrm{geo}^{(i, j)}\right|^2}{\Delta_\mathrm{geo}^{(i, j)} + \beta}.
    \label{eq:geodesicloss}
\end{equation}

Note how the proposed loss function in Eq.~(\ref{eq:geodesicloss}) makes use of the fact that all low-dimensional representations $\mathbf z^{(l)}$ are now readily available thanks to the new batch-wise training architecture from Sec.~\ref{sec:trainingarchitecture}.
Naively replacing $\mathcal L_\mathrm{siam}$ with $\mathcal L_\mathrm{geo}$ without making changes to the training procedure produces gravely inadequate channel charts.
We suspect that this is due to the high number of degrees of freedom in Eq.~(\ref{eq:geodesicloss}), where $\rho_\mathrm{geo}^{(i, j)}$ depends on many channel chart positions $\mathbf z^{(l)}$ if $M^{(i, j)}$ is large.
This causes the optimization algorithm to run into local optima early during training.
To mitigate this issue, we introduce path sub-sampling, where $\rho_\mathrm{geo}$ is approximated by a path that skips over datapoints along $q^{(i, j)}_m$, taking $s$ hops at a time.
We call $s$ the sub-sampling factor and calculate the length of the sub-sampled geodesic path through the chart as
\[
    \rho_\mathrm{geo}^{(i, j \vert s)} = \sum_{m=1}^{\left\lceil M^{(i, j)} / s \right\rceil} \left\lVert\mathbf z^{\left(q^{(i, j)}_{s (m-1)}\right)} - \mathbf z^{\left(q^{(i, j)}_{\min \left\{sm, M^{(i, j)}\right\}}\right)} \right\rVert
\]
and replace $\rho_\mathrm{geo}^{(i, j)}$ in Eq.~(\ref{eq:geodesicloss}) by $\rho_\mathrm{geo}^{(i, j \vert s)}$.
The sub-sampling factor $s$ is adjusted over the training steps and per path to ensure convergence towards the global optimum.
In the first training steps, one might chose $s = M^{(i, j)}$ so that $\mathcal L_\mathrm{geo}$ becomes identical to $\mathcal L_\mathrm{siam}$, which is helpful to achieve some initial convergence in a randomly initialized neural network.
Over the training steps, $s$ is decreased, for example based on a desired maximum length of the individual path segments.
The adjustment of $s$ over the training steps is needed only to ensure convergence towards the global optimum.
Finding the ideal adjustment scheme remains a topic for future work, and has to be considered in conjunction with other hyperparameters such as the learning rate.

\section{Third Improvement: Uncertainty Modeling}
\label{sec:uncertainty}
\subsection{Shortcoming of Current Approach}
A Gaussian log-likelihood function can be expressed as 
\[
    \mathcal L_\mathrm{gauss} = \mathrm{const.} -\frac{|x-\mu|^2}{2\sigma^2},
\]
where ``const.'' is some term that is constant with respect to $x$, and $\mu$ and $\sigma^2$ are the mean and variance of the normal distribution.
Comparing $\mathcal L_\mathrm{gauss}$ to $\mathcal L_\mathrm{siam}$ or $\mathcal L_\mathrm{geo}$ reveals that these loss functions can be viewed as Gaussian log-likelihood functions, with $\mu = \Delta_\mathrm{geo}^{(i,j)}$ and $\sigma^2 = \frac{1}{2} \left(\Delta_\mathrm{geo}^{(i,j)} + \beta\right)$.
In this interpretation, the modeled variance of the distribution (and hence, more abstractly, the uncertainty we have about $\Delta_\mathrm{geo}^{(i, j)}$) is proportional to $\Delta_\mathrm{geo}^{(i,j)} + \beta$.
Intuitively, this makes sense: For large path lengths $\Delta_\mathrm{geo}^{(i,j)}$, we have less certainty about the exact length of the path.
Our uncertainty decreases for shorter path lengths down to some lower limit, modeled by $\beta$.
Having said this, we may sometimes have additional information related to the geodesic path, which may allow us to deduce a better estimate for $\sigma^2$.
The intuition-based choice of $\sigma^2$, as in $\mathcal L_\mathrm{siam}$ and $\mathcal L_\mathrm{geo}$ is evidently sub-optimal and does not permit knowledge of uncertainties to be taken into account.

\subsection{Suggested Solution}
\begin{figure}
    \centering
	\begin{tikzpicture}
		\begin{axis}[
				width=0.8\columnwidth,
				height=0.5\columnwidth,	
				colorbar,
				colorbar style={
					scaled y ticks=false,
					ylabel={},
					width=0.05*\pgfkeysvalueof{/pgfplots/parent axis width},
					ticklabel style={/pgf/number format/fixed},
					xshift = -0.1cm
				},
				colormap/viridis,
				point meta max=0.03,
				point meta min=0,
				tick align=outside,
				tick pos=left,
				xlabel={ADP Dissimilarity \(\displaystyle \Delta_\mathrm{ADP}\)},
				xmin=0, xmax=25,
				xtick style={color=black},
				ylabel={Euc. Distance \(\displaystyle d\) [m]},
				ymin=0, ymax=0.12,
				yticklabel style={
						/pgf/number format/fixed,
						/pgf/number format/precision=5
				},
				scaled y ticks=false,
				ytick style={color=black}
			]
				\addplot graphics [includegraphics cmd=\pgfimage,xmin=0, xmax=25, ymin=0, ymax=0.12] {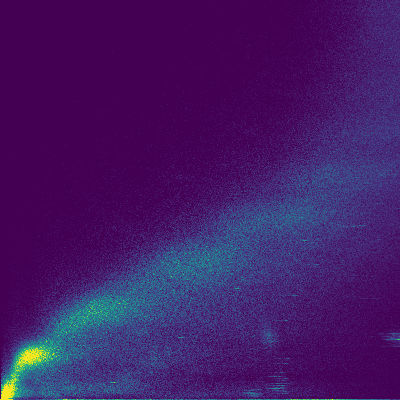};
				\node [fill = white, opacity = 1.0, fill opacity = 0.7, text opacity = 1.0] at (axis cs: 8, 0.1) {$p(d \,\vert\, \Delta_\mathrm{ADP})$};
		\end{axis}
	\end{tikzpicture}
	\vspace{-0.3cm}
    \caption{Conditional distribution $p(d \,\vert\, \Delta_\mathrm{ADP})$, empirically determined from all distinct pairs $(i,j) \in \left\{ 1,\ldots,L \right\}^2$}
    \label{fig:d_given_delta_adp}
	\vspace{-0.3cm}
\end{figure}

To develop an uncertainty-aware framework for Channel Charting, we model the dissimilarities $\Delta_\mathrm{time}$, $\Delta_\mathrm{ADP}$ and the physical Euclidean distances $d^{(i, j)} = \lVert \mathbf x^{(i)} - \mathbf x^{(j)} \rVert$ as random variables.
Considering all possible distinct datapoint pairs $(i, j)$, we observe $\frac{L^2 - L}{2}$ realizations of these random variables.

\subsubsection{Conditional Distributions $p(d \,\vert\, \Delta)$}
We are particularly interested in deriving a model for the conditional distributions of $d$ given a dissimilarity metric, such as $p(d \,\vert\, \Delta_\mathrm{ADP})$:
Having such a model will provide an estimated distribution of the datapoint pair distance $d^{(i, j)}$ from a dissimilarity, which is easily computed from the dataset.
A section of the empirically observed conditional distribution $p(d \,\vert\, \Delta_\mathrm{ADP})$ on dataset $\mathcal S$ is shown in Fig.~\ref{fig:d_given_delta_adp}.
$p(d \,\vert\, \Delta_\mathrm{ADP})$ is dependent on the number of antennas $B$, $M_\mathrm{row}$ and $M_\mathrm{col}$ in the system and we suspect that it is also affected by the placement of the antennas and possibly even by the type of environment.
In practice, we must assume that we do not have access to the empirical joint distribution $p(d \,\vert\, \Delta_\mathrm{ADP})$ (like in Fig.~\ref{fig:d_given_delta_adp}), since it was found using the ``ground truth'' distances $d$, which are assumed to be unknown.
Therefore, we need to manually specify models for $p(d \,\vert\, \Delta)$, for all dissimilarity metrics $\Delta$.

\subsubsection{Gaussian Approximation for $p(d \,\vert\, \Delta)$}
Practical considerations impose limitations on the model for $p(d \,\vert\, \Delta)$, namely:
\begin{itemize}
    \item The model should rely on as few hyperparameters as possible so that only few tweaks are necessary to adapt the system to a different environment and
    \item the model should facilitate the efficient computation of the distribution of sums of distances.
\end{itemize}
As a pragmatic compromise, we make the assumption that $d \,\vert\, \Delta \sim \mathcal N(\mu, \sigma^2)$, where $\mu$ and $\sigma^2$ are a function of $\Delta$.
We acknowledge that this model oversimplifies $p(d \,\vert\, \Delta)$ in many cases and has shortcomings, including nonzero probability densities for $d < 0$.
However, it is this simplification that makes the uncertainty-based framework for Channel Charting computationally tractable and practically feasible.
In the context of this work, we do not expand on the calculation of $\mu$ and $\sigma^2$ from the dissimilarity $\Delta$ and only note that we deduced a simple heuristic for the implementation.

\subsubsection{Shortest Path Algorithm}
Now that we have a model for the distribution of pairwise distances $d^{(i,j)} \,\vert\, \Delta^{(i, j)} $, we can compute the distribution of geodesic distances $d_\mathrm{geo}^{(i, j)}$.
In essence, we want to replicate the computation of geodesic dissimilarities $\Delta_\mathrm{geo}$ as in Sec.~\ref{sec:geodesic}, except that we now have a fully connected graph $G$ where the edge weights $d^{(i, j)} \,\vert\, \Delta$ are no longer deterministic.
This poses several challenges: First, the need to compute a $k$-neighbors graph for nondeterministic edge weights, running a shortest path algorithm on said $k$-neighbors graph and, once the shortest paths $q^{(i, j)}_m$ have been determined, the computation of the distributions of $d_\mathrm{geo}^{(i, j)}$ as the sum of nondeterministic edge weights along the paths.

Again, we need to make some pragmatic assumptions to keep the problem computationally tractable.
The first simplification is based on the realization that the paths $q^{(i, j)}_m$ do not necessarily need to be the very shortest paths through the manifold anymore, now that $\mathcal L_\mathrm{geo}$ as in Sec.~\ref{sec:geodesic} compares path lengths through the channel chart to path lengths through the manifold.
This is different than with $\mathcal L_\mathrm{siam}$, where the Euclidean distance in the Channel Chart was used and it was very important for the paths to be as straight as possible to match the linear, direct path of the Euclidean distance.
Nevertheless, even with $\mathcal L_\mathrm{geo}$, the paths should still be short, since $q^{(i, j)}_m$ may be sub-sampled as previously described.
Therefore, we suggest the following approach for finding \emph{short} paths (not necessarily short\emph{est} paths) through the graph $G$:
\begin{itemize}
    \item Generate a realization of $d^{(i, j)} \,|\, \Delta$ for every type of dissimilarity metric $\Delta$ and every distinct pair / edge $(i, j)$.
    \item Construct the fully connected graph $G$ with the realizations of $d^{(i, j)} \,|\, \Delta$ as edge weights. If there are multiple dissimilarity metrics, pick the smallest realization $d^{(i, j)}$ for each edge and remember this choice for later.
    \item As previously, run the shortest path algorithm on the $k$-neighbors graph of $G$ to find the shortest paths $q^{(i, j)}_m$.
    \item Repeat this procedure several times and remember all $q^{(i, j)}_m$. Each time, the shortest paths for one particular realization are found, which are short paths when considering a graph with probabilistic edge weights.
\end{itemize}

Now that we have a collection of short paths $q^{(i, j)}_m$, we can randomly select a batch of short paths in every training step.

\subsubsection{Total Path Lengths}
$\Delta_\mathrm{geo}^{(i,j)}$ is the sum of individual dissimilarities along a path $q^{(i, j)}_m$.
In the same way, $d_\mathrm{geo}^{(i,j)}$ models the distribution of the sum of individual distances $d^{(i, j)}$ along the path, that is
\[
    d_{\mathrm{geo}}^{(i,j)} = \sum_{m=1}^{M^{(i, j)}} d^{\left(q^{(i, j)}_{m-1}, q^{(i, j)}_m\right)} \sim \mathcal N\left(\mu_\mathrm{geo}^{(i, j)}, \left(\sigma_\mathrm{geo}^{(i, j)}\right)^2\right).
\]
By modeling all conditional distributions $p(d \,\vert\, \Delta_\mathrm{ADP})$ as Gaussian, we can be sure that the total length $d_\mathrm{geo}^{(i, j)}$ of some path $q^{(i, j)}_m$ is also normally distributed in our model.
$\mu_\mathrm{geo}^{(i, j)}$ is easily determined as the sum of all means along the path, i.e.,
\[
    \mu_{\mathrm{geo}}^{(i,j)} = \sum_{m=1}^{M^{(i, j)}} \mu^{\left(q^{(i, j)}_{m-1}, q^{(i, j)}_m\right)}
\]
where $\mu^{\left(q^{(i, j)}_{m-1}, q^{(i, j)}_m\right)}$ is the modeled mean of the distribution of $d^{(i, j)} \,\vert\, \Delta$ for the remembered choice of dissimilarity metric.
Computing the variance of $d_\mathrm{geo}^{(i,j)}$ is not as straightforward though, since it requires a model for the correlation between different dissimilarities to determine the covariances, i.e.,
\[
    \left(\sigma_{\mathrm{geo}}^{(i,j)}\right)^2 = \sum_{m=1}^{M_{i, j}} \left(\sigma^{\left(q^{(i, j)}_{m-1}, q^{(i, j)}_m\right)}\right)^2 + 2 \sum_{1 \leq a < b \leq m} \text{Cov}\left(d_a, d_b\right)
\]
where we used abbreviations $d_a = d^{\left(q^{(i, j)}_{a-1}, q^{(i, j)}_a\right)}$ and $d_b = d^{\left(q^{(i, j)}_{b-1}, q^{(i, j)}_b\right)}$.
Remember that $d_a$ and $d_b$ are Gaussian probability distributions whose distribution parameters were derived from some dissimilarity metric $\Delta$ based on a conditional distribution $p(d\,\vert\,\Delta)$.
$d_a$ and $d_b$ may have been derived from the same dissimilarity metric (e.g., $d_a \,\vert\,\Delta_\mathrm{ADP}$ and $d_b \,\vert\,\Delta_\mathrm{ADP}$) or from different metrics (e.g., $d_a \,\vert\,\Delta_\mathrm{ADP}$ and $d_b \,\vert\,\Delta_\mathrm{time}$), which will affect their correlation properties.
Again, to compute $\text{Cov}\left(d_a, d_b\right)$, we make some reasonable, but imperfect simplifications to ensure computational feasibility:
\begin{itemize}
    \item If the distributions of $d_a$ and $d_b$ were derived from different dissimilarity metrics, we assume $\text{Cov}\left(d_a, d_b\right) = 0$.
    \item We set $\text{Cov}\left(d_a, d_b\right) = 0$ if $d_a$, $d_b$ are derived from $\Delta_\mathrm{ADP}$.
    \item If $d_a$ and $d_b$ were derived from $\Delta_\mathrm{time}$, we assume $\left(\text{Cov}\left(d_a, d_b\right)\right)^2 = \text{Var}(d_a) \text{Var}(d_b)$, i.e., perfectly correlated random variables. This is the worst-case assumption in the sense that it maximizes the modeled uncertainty $\sigma_\mathrm{geo}^{(i, j)}$. This choice is reasonable if the \ac{UE} speed is unknown, in which case $d_a\,\vert\,\Delta_\mathrm{time}$ and $d_b\,\vert\,\Delta_\mathrm{time}$ may be perfectly correlated if the speed is constant.
    \item Keep in mind that the correlation model for distance distributions derived from $\Delta_\mathrm{time}$ may change if (noisy) speed measurements are available as in \cite{stahlke2023velocity}.
\end{itemize}

\subsubsection{Uncertainty-aware loss function}
Drawing lessons from Sec.~\ref{sec:geodesic} related to the geodesic loss, but now also incorporating the Gaussian uncertainty model for $d_\mathrm{geo}$, we formulate a new uncertainty-aware geodesic loss function
\begin{equation}
    \mathcal L_\mathrm{geo,unc} = \sum_{(i,j) \in \mathcal B} \frac{\left| \rho_\mathrm{geo}^{(i, j)} - \mu_\mathrm{geo}^{(i, j)}\right|^2}{2 \left(\sigma_\mathrm{geo}^{(i, j)}\right)^2},
    \label{eq:uncertaintyloss}
\end{equation}
where $\rho_\mathrm{geo}^{(i, j)}$ denotes the length of the (potentially sub-sampled) geodesic path through the channel chart as previously.

\section{Evaluation}
\label{sec:evaluation}

\begin{figure}
    \centering
    \begin{subfigure}{0.49\columnwidth}
        \begin{tikzpicture}
            \begin{axis}[
                width=0.7\textwidth,
                height=0.7\textwidth,
                scale only axis,
                enlargelimits=false,
                axis on top,
                xlabel = {Coordinate $\tilde z_1 ~ [\mathrm{m}]$},
                ylabel = {Coordinate $\tilde z_2 ~ [\mathrm{m}]$},
                ylabel shift = -8 pt,
                xlabel shift = -4 pt
            ]
                \addplot graphics[xmin=-13.028850841408582, xmax=3.6687306000188045, ymin=-15.031928608687338, ymax=-1.0398224119322146] {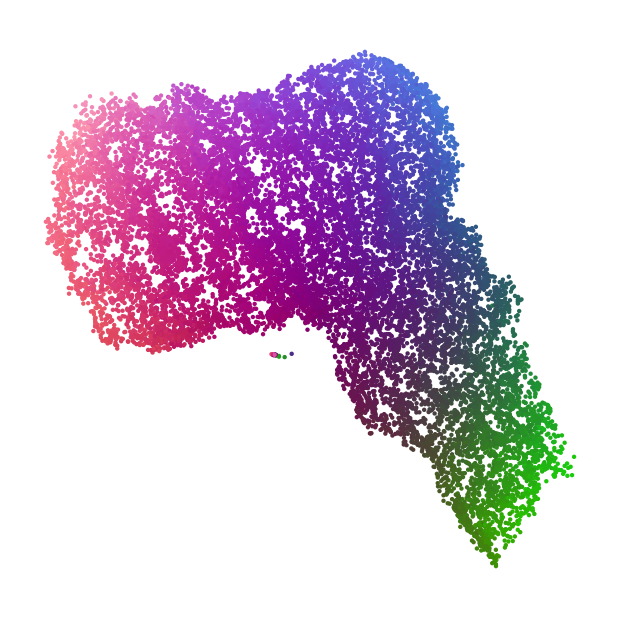};
            \end{axis}
        \end{tikzpicture}
    \end{subfigure}
    \begin{subfigure}{0.49\columnwidth}
    	\begin{tikzpicture}
    		\begin{axis}[
    			width=0.7\textwidth,
    			height=0.7\textwidth,
    			scale only axis,
    			xmin=-13,
    			xmax=3.2,
    			ymin=-14.8,
    			ymax=-1.2,
    			xlabel = {Coordinate $\tilde z_1 ~ [\mathrm{m}]$},
    			ylabel = {Coordinate $\tilde z_2 ~ [\mathrm{m}]$},
    			ylabel shift = -8 pt,
    			xlabel shift = -4 pt,
    			xtick={-10, -6, -2, 2}
    		]
    			\addplot[thick,blue] graphics[xmin=-13,ymin=-14.8,xmax=3.2,ymax=-1.2] {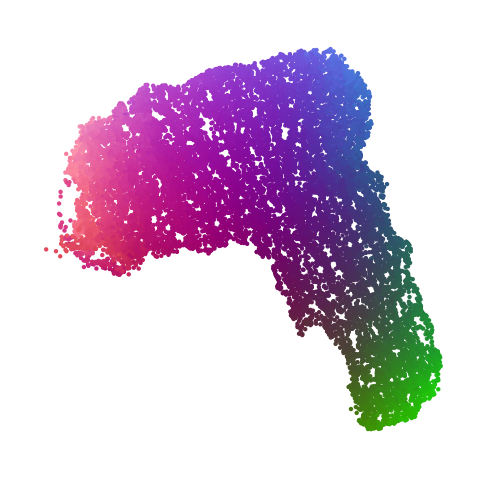};
    		\end{axis}
    	\end{tikzpicture}
    \end{subfigure}
    \vspace{-0.5cm}
    \caption{Learned Channel Chart trained on \ac{CSI} dataset $\mathcal S$ containing \acp{CIR} from all four antenna arrays using $\mathcal L_\mathrm{siam}$ and without any of the suggested improvements (left) and with loss function $\mathcal L_\mathrm{geo,unc} + \mathcal L_\mathrm{acc}$, i.e., with all suggested improvements (right), both after optimal affine transform $\mathcal T$.}
    \label{fig:cc_allarrays_transformed}
    \vspace{-0.3cm}
\end{figure}

\begin{figure}
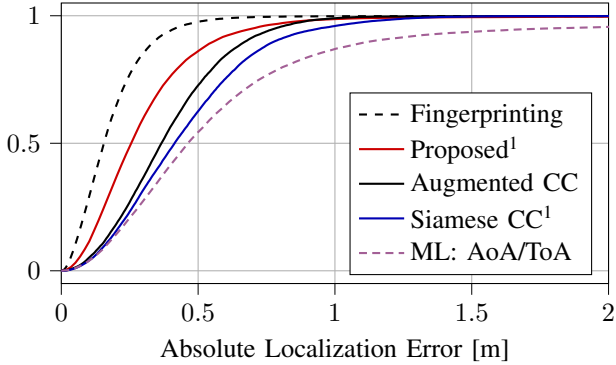

    \centering
    \include{fig/cdf}
    \vspace{-0.9cm}
    \textsuperscript{1}\scriptsize{Evaluated after optimal affine transform}
    \vspace{-0.2cm}
    \caption{Empirical cumulative distribution functions of absolute localization errors for classical (maximum likelihood) and Channel Charting methods. Baselines taken from \cite{asilomar2023}, ``Fingerprinting'' is supervised training of same \ac{NN}.}
    \label{fig:cdf}
    \vspace{-0.4cm}
\end{figure}

In our evaluation, we focus on Channel Charting results that were obtained after applying \emph{all} suggested improvements, i.e., when training with the new batch-wise loss architecture from Sec.~\ref{sec:trainingarchitecture} and using loss function $\mathcal L_\mathrm{batch} = \mathcal L_\mathrm{geo-unc} + \mathcal L_\mathrm{acc}$.
We also dedicate a few paragraphs to an ablation study that determines the contribution of individual changes in Sec.~\ref{sec:ablation}, but acknowledge that the importance individual contributions greatly depends on the scenario and hyperparameters.
We evaluate the \ac{FCF} on the training set itself, which is a reasonable evaluation method for self-supervised training.
We found similar performance when evaluating on a separate test set.

\begin{table*}
    \centering
    \captionof{table}{Localization performance comparison, training and evaluation on $\mathcal S$: CSI data from all four antenna arrays}
    \vspace{-0.1cm}
    \begin{tabular}{r c | c c c c c c}
        & \textbf{Loss} & \textbf{MAE $\downarrow$} & \textbf{DRMS $\downarrow$} & \textbf{CEP $\downarrow$} & \textbf{R95 $\downarrow$} & \textbf{KS $\downarrow$} & \textbf{CT/TW $\uparrow$} \\ \hline
        Classical Max. Likelihood: ToA + AoA\footnotemark[2] \cite{asilomar2023} & $\mathcal L_\mathrm{AoA/ToA}$ & $0.676\,\mathrm m$ & $1.228\,\mathrm m$ & $0.462\,\mathrm m$ & $1.763\,\mathrm m$ & $0.214$ & $0.965/0.970$ \\
        Siamese CC\footnotemark[1] \cite{stephan2023angle} & $\mathcal L_\mathrm{siam}$ & $0.490\,\mathrm m$ & $0.584\,\mathrm m$ & $0.441\,\mathrm m$ & $1.026\,\mathrm m$ & $0.071$ & $0.996/0.996$ \\
        Augmented CC: Classical + CC\footnotemark[2] \cite{asilomar2023} & $\mathcal L_\mathrm{aug}$ & $0.401\,\mathrm m$ & $0.483\,\mathrm m$ & $0.369\,\mathrm m$ & $0.789\,\mathrm m$ &
        $0.070$ & $0.995/0.995$ \\
		Proposed\footnotemark[1] (this paper, all improvements) & $\mathcal L_\mathrm{geo,unc} + \mathcal L_\mathrm{acc}$ & $\mathbf{0.330\,\mathrm{m}}$ & $\mathbf{0.409\,\mathrm{m}}$ & $\mathbf{0.277\,\mathrm{m}}$ & $\mathbf{0.719\,\mathrm{m}}$ & $\mathbf{0.069}$ & $\mathbf{0.998/0.998}$
    \end{tabular}

    \vspace{0.5em}
    \textsuperscript{1}\scriptsize{MAE / DRMS / CEP / R95 evaluated after optimal affine transform}

    \textsuperscript{2}\scriptsize{See \cite{asilomar2023} for definition of loss functions $\mathcal L_\mathrm{AoA/ToA}$ (classical triangulation / multilateration, no \ac{NN}) and $\mathcal L_\mathrm{aug}$ (combination of classical methods with Channel Charting)}
    \label{tab:performance}
    \vspace{-0.4cm}
\end{table*}

Note that performance can vary across training iterations, and that, depending on the initialization, the neural network may sometimes not converge to the global optimum, which will be apparent from the loss.
Therefore, the performance metrics are averaged over 20 training passes, possibly omitting results with unreasonably high loss from the average.

\subsection{Optimal Affine Transform}
While the \ac{FCF} should preserve relative positions, the channel chart's coordinate frame does not need to match the physical coordinate frame.
Even if the all distances between datapoints are correctly predicted in a physically meaningful unit, the channel chart is only unique up to a rigid transformation.
We take inspiration from \cite{fraunhofer_cc} and evaluate the channel chart after a transform $\mathcal T$ to the physical coordinate frame.
To find $\mathcal T(\mathbf x) = \hat { \mathbf A } \mathbf x + \hat { \mathbf b }$, we solve the least squares problem
\begin{equation}
    (\mathbf{\hat A}, \mathbf{\hat b}) = \argmin\limits_{(\mathbf{A}, \mathbf{b})} \sum_{l = 1}^L \lVert\mathbf{A} {\mathbf z }^{(l)} + \mathbf b - \mathbf x^{(l)} \rVert^2,
    \label{eq:affinetransform}
\end{equation}
where $\mathbf z^{(l)} = \mathcal C_\theta\left(\mathbf H^{(l)}\right)$ denotes the position estimate obtained from the \ac{FCF}.
While the ground truth positions $\mathbf x^{(l)}$, which are necessary to apply Eq. (\ref{eq:affinetransform}), are assumed to be unavailable, the results from \cite{asilomar2023} and \cite{pihlajasalo2020absolute} indicate that similar or better absolute localization can be achieved by taking information like angle / time of arrival into account during training.

\subsection{Evaluation Metrics}
We obtain absolute position estimates $\hat { \mathbf x }^{(l)} = \mathcal T_\mathrm{c} \circ \mathcal C \left(\mathbf H^{(l)} \right)$, where $\mathcal T \circ \mathcal C_\theta$ denotes the composition of $\mathcal T$ and $\mathcal C_\theta$, and borrow performance metrics from radio localization literature, which rely on the distance error $e^{(l)} = \left\lVert \mathbf x^{(l)} - \hat {\mathbf x}^{(l)} \right\rVert$, to quantify the accuracy of $\hat { \mathbf x }$.
The \ac{MAE} and \ac{DRMS} are defined as $\mathrm{MAE} = \frac{1}{L} \sum_{l=1}^L e^{(l)}$ and $\mathrm{DRMS} = \sqrt{\frac{1}{L} \sum_{l=1}^L \left(e^{(l)}\right)^2}$.
Furtheremore, we define the \ac{CEP} and the \ac{R95} as the median and 95\textsuperscript{th} percentile of the empirical distribution of $e^{(l)}$.
We adopt the dimensionality reduction performance metrics \ac{CT} and \ac{TW} (level of preservation of local geometry), and \ac{KS} (level of preservation of global geometry) from Channel Charting literature, as defined, e.g., in \cite{fraunhofer_cc} or \cite{stephan2023angle}.

\subsection{Multi-Array Results}
After training with loss function $\mathcal L_\mathrm{batch} = \mathcal L_\mathrm{geo,unc} + \mathcal L_\mathrm{acc}$ on the complete training set $\mathcal S$, we obtain the channel chart in Fig.~\ref{fig:cc_allarrays_transformed} (right).
Compared to the result without any of the suggested improvements, the overall global shape is now captured more accurately, including the $90^\circ$ angle in the corner.
Overall localization performance is greatly improved as is apparent from the observed empirical \ac{CDF} in Fig.~\ref{fig:cdf} and from the performance metrics in Tab.~\ref{tab:performance}.

\subsection{Single-Array Results}
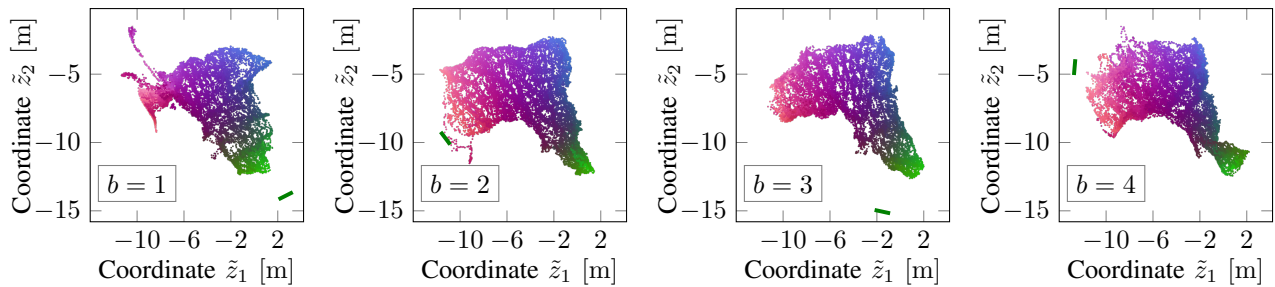
\begin{figure*}
    \centering
    \begin{subfigure}{0.23\textwidth}
        \centering
		\begin{tikzpicture}
			\begin{axis}[
				width=0.68\textwidth,
				height=0.68\textwidth,
				scale only axis,
				xmin=-14,
				xmax=4.2,
				ymin=-15.8,
				ymax=-0.2,
				xlabel = {Coordinate $\tilde z_1 ~ [\mathrm{m}]$},
				ylabel = {Coordinate $\tilde z_2 ~ [\mathrm{m}]$},
				ylabel shift = -8 pt,
				xlabel shift = -4 pt,
				xtick={-10, -6, -2, 2}
			]
				\addplot[thick,blue] graphics[xmin=-13,ymin=-14.8,xmax=3.2,ymax=-1.2] {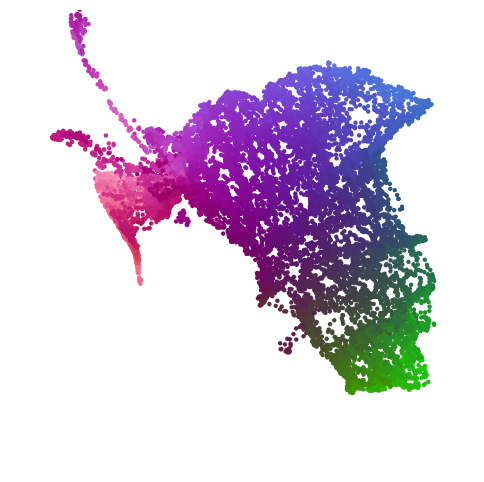};
                \node (arraya) at (axis cs: 2.6747, -13.8973) [fill = green!50!black, minimum width = 1.5, minimum height = 6, inner sep = 0pt, rotate = 116.8, anchor = center] {}; 
				\node [fill = white, draw = black!50!white, opacity = 1.0, fill opacity = 0.7, text opacity = 1.0] at (axis cs: -10, -13) {$b = 1$};
			\end{axis}
		\end{tikzpicture}
    \end{subfigure}
    \begin{subfigure}{0.23\textwidth}
        \centering
    	\begin{tikzpicture}
			\begin{axis}[
				width=0.68\textwidth,
				height=0.68\textwidth,
				scale only axis,
				xmin=-14,
				xmax=4.2,
				ymin=-15.8,
				ymax=-0.2,
				xlabel = {Coordinate $\tilde z_1 ~ [\mathrm{m}]$},
				ylabel = {Coordinate $\tilde z_2 ~ [\mathrm{m}]$},
				ylabel shift = -8 pt,
				xlabel shift = -4 pt,
				xtick={-10, -6, -2, 2}
			]
				\addplot[thick,blue] graphics[xmin=-13,ymin=-14.8,xmax=3.2,ymax=-1.2] {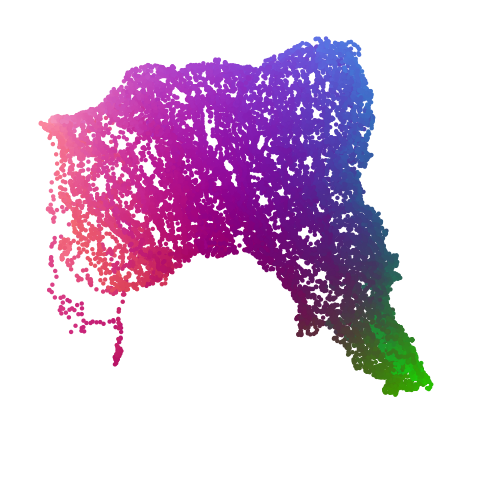};
                \node (arrayb) at (axis cs: -11.250275, -9.689025) [fill = green!50!black, minimum width = 1.5, minimum height = 6, inner sep = 0pt, rotate = 37.2, anchor = center] {}; 
				\node [fill = white, draw = black!50!white, opacity = 1.0, fill opacity = 0.7, text opacity = 1.0] at (axis cs: -10, -13) {$b = 2$};
			\end{axis}
		\end{tikzpicture}
    \end{subfigure}
    \begin{subfigure}{0.23\textwidth}
        \centering
		\begin{tikzpicture}
			\begin{axis}[
				width=0.68\textwidth,
				height=0.68\textwidth,
				scale only axis,
				xmin=-14,
				xmax=4.2,
				ymin=-15.8,
				ymax=-0.2,
				xlabel = {Coordinate $\tilde z_1 ~ [\mathrm{m}]$},
				ylabel = {Coordinate $\tilde z_2 ~ [\mathrm{m}]$},
				ylabel shift = -8 pt,
				xlabel shift = -4 pt,
				xtick={-10, -6, -2, 2}
			]
				\addplot[thick,blue] graphics[xmin=-13,ymin=-14.8,xmax=3.2,ymax=-1.2] {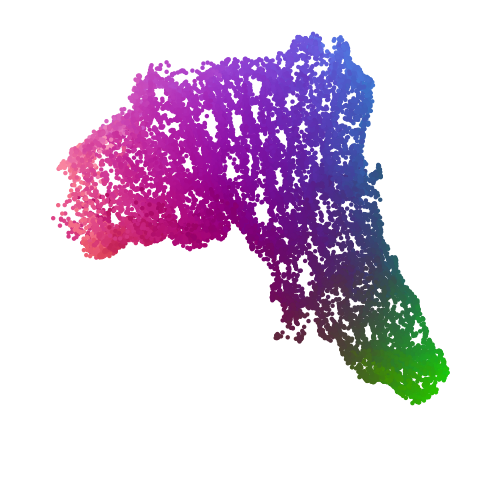};
                \node (arrayc) at (axis cs: -1.531375, -15.0595) [fill = green!50!black, minimum width = 1.5, minimum height = 6, inner sep = 0pt, rotate = 77.8, anchor = center] {}; 
				\node [fill = white, draw = black!50!white, opacity = 1.0, fill opacity = 0.7, text opacity = 1.0] at (axis cs: -10, -13) {$b = 3$};
			\end{axis}
		\end{tikzpicture}
    \end{subfigure}
    \begin{subfigure}{0.23\textwidth}
        \centering
		\begin{tikzpicture}
			\begin{axis}[
				width=0.68\textwidth,
				height=0.68\textwidth,
				scale only axis,
				xmin=-14,
				xmax=4.2,
				ymin=-15.8,
				ymax=-0.2,
				xlabel = {Coordinate $\tilde z_1 ~ [\mathrm{m}]$},
				ylabel = {Coordinate $\tilde z_2 ~ [\mathrm{m}]$},
				ylabel shift = -8 pt,
				xlabel shift = -4 pt,
				xtick={-10, -6, -2, 2}
			]
				\addplot[thick,blue] graphics[xmin=-13,ymin=-14.8,xmax=3.2,ymax=-1.2] {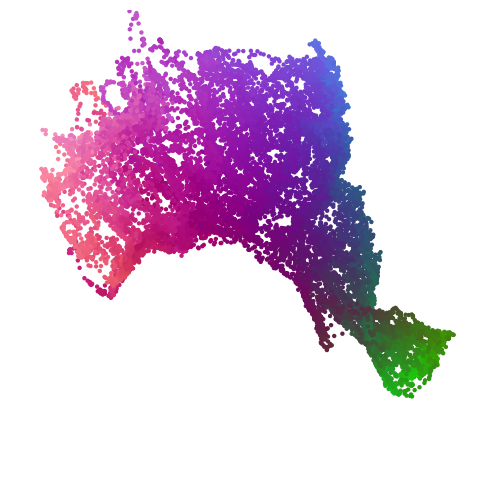};
                \node (arrayd) at (axis cs: -12.684425, -4.483325) [fill = green!50!black, minimum width = 1.5, minimum height = 6, inner sep = 0pt, rotate = -4.87, anchor = center] {}; 
				\node [fill = white, draw = black!50!white, opacity = 1.0, fill opacity = 0.7, text opacity = 1.0] at (axis cs: -10, -13) {$b = 4$};
			\end{axis}
		\end{tikzpicture}
    \end{subfigure}
    \vspace{-0.3cm}
    \caption{Exemplary channel charts (after optimal affine transform), trained only on CSI data from individual antenna arrays ($\mathcal S|_{b=1}$, $\mathcal S|_{b=2}$, $\mathcal S|_{b=3}$ and $\mathcal S|_{b=4}$)}
    \label{fig:cc_singleant}
    \vspace{-0.4cm}
\end{figure*}

The original paper on Channel Charting \cite{studer_cc} already claimed the feasibility of the scheme under \ac{NLoS} conditions.
While this assertion has been confirmed several times \cite{stephan2023angle, stahlke2023velocity}, we want to investigate the particularly hard problem of \ac{NLoS} localization when channel observations are only available from the perspective of a single base station site.
To this end, we train and evaluate the \ac{FCF} on $\mathcal S|_{b=\tilde b}$, which contains only \ac{CSI} from individual antenna arrays.
From the resulting channel charts in Fig.~\ref{fig:cc_singleant}, we see that the localization performance in the \ac{NLoS} areas (around the corner / behind the scatterer) is surprisingly good.
Tab.~\ref{tab:performance-singlearray} also shows the significant improvement over $\mathcal L_\mathrm{siam}$ in this regard.
A common error is that the optimizer does not find the global minimum of the loss function.
The effect is visible in Fig.~\ref{fig:cc_singleant} for $b=4$: The channel chart looks ``twisted'' in the \ac{NLoS} area.

\subsection{Ablation Study}
\label{sec:ablation}

\begin{table}
    \centering
    \captionof{table}{Performance for selectively applied loss enhancements}
    \vspace{-0.2cm}
    \begin{tabular}{r | c | c c}
        \textbf{Dataset} & \textbf{Loss} $\mathcal L_\mathrm{batch}$ & \textbf{MAE}\footnotemark[1] & \textbf{CEP}\footnotemark[1] \\ \hline
        $\mathcal S$ & $\mathcal L_\mathrm{geo}$ & $0.340\,\mathrm{m}$ & $0.310\,\mathrm{m}$ \\
        $\mathcal S$ & $\mathcal L_\mathrm{geo} + \mathcal L_\mathrm{acc}$ & $0.318\,\mathrm{m}$ & $0.291\,\mathrm{m}$ \\
        $\mathcal S$ & $\mathcal L_\mathrm{geo,unc}$ & $0.339\,\mathrm{m}$ & $0.282\,\mathrm{m}$ \\
        $\mathcal S_{b=3}$ & $\mathcal L_\mathrm{geo}$ & $0.703\,\mathrm{m}$ & $0.614\,\mathrm{m}$ \\
        $\mathcal S_{b=3}$ & $\mathcal L_\mathrm{geo} + \mathcal L_\mathrm{acc}$ & $0.614\,\mathrm{m}$ & $0.528\,\mathrm{m}$ \\
        $\mathcal S_{b=3}$ & $\mathcal L_\mathrm{geo,unc}$ & $0.560\,\mathrm{m}$ & $0.460\,\mathrm{m}$ \\
    \end{tabular}
    \label{tab:ablation_study}

    \textsuperscript{1}\scriptsize{MAE and CEP evaluated after optimal affine transform}
    \vspace{-0.3cm}
\end{table}

To determine the contribution of the three individual changes from Sections~\ref{sec:acceleration}, \ref{sec:geodesic} and \ref{sec:uncertainty} to the overall improvement, we apply the changes to the loss function selectively.
While not an exhaustive overview over all possible combinations, Tab.~\ref{tab:ablation_study} lists the resulting \acp{MAE} and \acp{CEP} for various loss functions and for an \ac{FCF} that is either trained and evaluated on the full dataset $\mathcal S$, or on the dataset $\mathcal S|_{b = 3}$.
It is important to keep in mind that the performance of a particular loss function not only depends on the construction of the loss function, but also on the choice of hyperparameters, which may be further tuned for other losses.
The results in Tab.~\ref{tab:ablation_study} can therefore interpreted as indiciations for the achievable performance, but not as definitive results.

Surprisingly, when trained on the complete dataset $\mathcal S$, we find that the loss function $\mathcal L_\mathrm{batch} = \mathcal L_\mathrm{geo} + \mathcal L_\mathrm{acc}$ performs roughly equally well as $\mathcal L_\mathrm{batch} = \mathcal L_\mathrm{geo, unc} + \mathcal L_\mathrm{acc}$.
This is possibly explained by unprecise models $p(d \,\vert\, \Delta_\mathrm{ADP})$ and $p(d \,\vert\, \Delta_\mathrm{time})$, sub-optimal correlation models, or maybe we gained little information by modeling uncertainties.
On the other hand, when trained on only $\mathcal S\vert_{b=3}$, the uncertainty-aware framework with loss function $\mathcal L_\mathrm{batch} = \mathcal L_\mathrm{geo, unc}$ or $\mathcal L_\mathrm{batch} = \mathcal L_\mathrm{geo, unc} + \mathcal L_\mathrm{acc}$ clearly improves performance.

\begin{table}
    \centering
    \captionof{table}{Performance for $\mathcal S|_{b = \tilde b}$: CSI data from a single antenna array}
    \vspace{-0.2cm}
	\begin{tabular}{r | c | c}
        \textbf{Ant. Array} & \textbf{MAE\textsuperscript{1}} for $\mathcal L_\mathrm{geo,unc} + \mathcal L_\mathrm{acc}$ & \textbf{MAE\textsuperscript{1}} for $\mathcal L_\mathrm{siam}$ \\ \hline
		$\tilde b = 1$ & $\approx 1.05\,\mathrm{m}$ & $\approx 2.04\,\mathrm{m}$ \\
		$\tilde b = 2$ & $\approx 0.48\,\mathrm{m}$ & $\approx 2.23\,\mathrm{m}$ \\
		$\tilde b = 3$ & $\approx 0.49\,\mathrm{m}$ & $\approx 1.37\,\mathrm{m}$ \\
		$\tilde b = 4$ & $\approx 0.73\,\mathrm{m}$ & $\approx 1.62\,\mathrm{m}$
	\end{tabular}

    \vspace{0.5em}
    \textsuperscript{1}\scriptsize{MAE evaluated after optimal affine transform}

    \label{tab:performance-singlearray}
    \vspace{-0.3cm}
\end{table}

\section{Conclusion and Outlook}
\label{sec:conclusion}

Our results show improvements over the state of the art, but we also introduce additional complexity to achieve them.
This likely hints at diminishing returns of further improving Channel Charting if existing techniques are already good.
Our suggestions also add hyperparameters, which require careful tuning.
Future work could focus on automatically deriving hyperparameters, especially those for the conditional distribution models $d\,\vert\,\Delta$.
For instance, if a model $d\,\vert\,\Delta_\mathrm{ADP}$ is known for one dissimilarity metric $\Delta_\mathrm{ADP}$, the measured dataset could be used to automatically derive similar models $d\,\vert\,\Delta$ for other dissimilarity metrics $\Delta$.
To test the robustness of our proposed methods, we also suggest applying them to other datasets.
We speculate that pre-training the \ac{FCF} on a free-space model for weight initialization could improve convergence to the global minimum, which may be another research direction.

\bibliographystyle{IEEEtran}
\bibliography{IEEEabrv,references}

\end{document}